\documentclass[amsmath,amssymb,aps,prd,showpacs,showkeys,nofootinbib]{revtex4}
\usepackage{graphicx}
\usepackage{dcolumn}
\usepackage{bm}
\usepackage{latexsym,color}
\begin{document}
\author{Kuantay Boshkayev}
\email{kuantay@mail.ru, kuantay@icra.it}
\affiliation{Physical and Technical Faculty of al-Farabi Kazakh National University,\\ 
al-Farabi ave. 71, 050040, Almaty, Kazakhstan.}
\title{What can we extract from quasi-periodic oscillations? }

\author{Donato Bini}
\email{binid@icra.it}
\affiliation{Dipartimento di Fisica, Universit\`a di Roma "Sapienza", Piazzale Aldo Moro 5, I-00185 Roma, Italy\\
ICRANet, Piazzale della Repubblica 10, I-65122 Pescara, Italy.}

\author{Jorge Rueda}
\email{jorge.rueda@icra.it}
\affiliation{Dipartimento di Fisica, Universit\`a di Roma "Sapienza", Piazzale Aldo Moro 5, I-00185 Roma, Italy\\
ICRANet, Piazzale della Repubblica 10, I-65122 Pescara, Italy.}

\author{Andrea Geralico}
\email{geralico@icra.it}
\affiliation{Dipartimento di Fisica, Universit\`a di Roma "Sapienza", Piazzale Aldo Moro 5, I-00185 Roma, Italy\\
ICRANet, Piazzale della Repubblica 10, I-65122 Pescara, Italy.}

\author{Marco Muccino}
\email{marco.muccino@icra.it}
\affiliation{Dipartimento di Fisica, Universit\`a di Roma "Sapienza", Piazzale Aldo Moro 5, I-00185 Roma, Italy\\
ICRANet, Piazzale della Repubblica 10, I-65122 Pescara, Italy.}

\author{Ivan Siutsou}
\email{siutsou@icranet.org}
\affiliation{Dipartimento di Fisica, Universit\`a di Roma "Sapienza", Piazzale Aldo Moro 5, I-00185 Roma, Italy\\
ICRANet, Piazzale della Repubblica 10, I-65122 Pescara, Italy.}


\begin{abstract}
In light of the relativistic precession model, we present here detailed analyses, extending the ones performed in the Schwarzschild and Kerr spacetimes. We consider the kilohertz quasi-periodic oscillations in the Hartle-Thorne spacetime, which describes the gravitational field of a rotating and deformed object. We derive the analytic formulas for the epicyclic frequencies in the Hartel-Thorne spacetime and by means of these frequencies we interpret the kilohertz quasi-periodic oscillations  of low-mass X-ray binaries of the atoll and Z - sources, on the basis of the relativistic precession model. Particularly we perform analyzes for Z -source: GX 5-1. We show that the quasi-periodic oscillations data can provide information on the parameters, namely, the mass, angular momentum and quadrupole moment of the compact objects in the low-mass X-ray binaries. 
\end{abstract}

\maketitle
\section{Introduction}
It is believed that the quasi-periodic oscillations (QPOs) data of the X-ray flux from low mass X-ray binaries (LMXRBs) may be used to test general relativity (GR) in the strong field regime \cite{ste-vie:1998, ste-vie:1999,  ste-etal:1999, ste-vie:2002}. QPOs appear in variabilities of several LMXRBs including those which contain a neutron star (NS). A certain kind of these oscillations, the so-called kilohertz (kHz) (or high-frequency) QPOs, come often in pairs with frequencies $f_L=f_{per}=f_{\phi}-f_r$, where $f_{\phi}$ is the azimutal (Keplerian) frequency, $f_{r}$ is the radial epicyclic frequency of the Keplerian motion, and $f_U=f_{\phi}=f_K$ is typically in the range $\sim50-1300\,\mathrm{Hz}$. This is of the same order as the range of frequencies characteristic for orbital motion close to a compact object. Accordingly, most kHz QPO models involve orbital motion in the inner regions of an accretion disk (see \cite{Kli:2006:CompStelX-Ray:,Lam-Bou:2007:ASSL:ShrtPerBS}).

In order to explain the QPOs, various types of models have been proposed. These are: (i) the Beat-frequency models, where one assumes that there is some "beating" of an orbital frequency by the spin frequency of the central object, (ii) the relativistic precession models, where the QPOs are associated with the orbital motion and the periastron or nodal precession of a particular orbit, (iii) the relativistic resonance models, where a type of resonance between the orbital and the epicyclic frequencies is assumed wherever they have simple integer ratios, and finally (iv) the preferred radii models, where some mechanism chooses a particular radius. These models generally assume the geodesic or almost geodesic orbits of the fluid elements in the accretion disc to be the source of the observed frequencies (see e.g. \cite{Kli:2006:CompStelX-Ray:}), while there are also models in which the frequencies are produced from oscillatory modes of the entire disc (see e.g. \cite{rez-etal:2003}). In one way or another all of these models use the properties of the orbits around the compact object onto which the accretion takes place. In our discussion we will refer to the models that assume that the QPOs are caused by the frequencies associated with the orbital motion of the material in the accretion disc such as the relativistic precession models (RPM)(see \cite{ste-vie:1999,2012MNRAS.422.2581P}).

The RPM has been proposed in a series of papers by \cite{ste-vie:1998,ste-vie:1999,ste-vie:2002}. It explains the kHz QPOs as a direct manifestation of modes of relativistic epicyclic motion of blobs arising at various radii $r$ in the inner parts of the accretion disk. The model identifies the lower and upper kHz QPOs with the periastron precession $f_{per}$ and Keplerian $f_K$ frequency.

In the past years, the RPM has been considered among the candidates for explaining the twin-peak QPOs in several LMXBs and related constraints on the sources have been discussed (see e.g., \citep{kar:1999,Zha-etal:2006:MONNR:kHzQPOFrCorr,bel-etal:2007a,Lam-Bou:2007:ASSL:ShrtPerBS,bar-bou:2008a,yan-etal:2009}). While some of the early works discuss these constraints in terms of both NS mass and spin and include also the NS oblateness \citep[][]{mor-ste:1999,ste-etal:1999}, most of the published implications for individual sources focus on the NS mass and neglect its rotation not to mention its oblateness.

In this paper, we perform analysis of mass estimate carried out by Stella \cite{ste-vie:1999} and Boutloukos et al \cite{bou-etal}. In particular, we consider rotating spacetimes that comprehend the effects of frame-dragging and quadrupole moment of the source and fit directly the correlation between the twin QPO frequencies. We show that good fits can be reached for the case of three parameters the mass, angular momentum and quadrupole moment rather than for the preferred combination of mass and angular momentum \cite{2010ApJ...714..748T}. The importance of the quadrupole moment has been emphasized  in several works \cite{1989GReGr..21.1047Q, 2009CQGra..26v5006B, 2012CQGra..29n5003B, 2012ApJ...756...82P, 2012MNRAS.422.2581P, bini2013}. Since the angular momentum of the source is in a non-trivial way related to its quadrupole moment, they should be considered together.


%
\section{The Hartle-Thorne metric}
The Hartle-Thorne metric describing the exterior field of a slowly rotating slightly deformed object is given by
\begin{equation}\label{ht1}
\begin{split}
ds^2=-\left(1-\frac{2{ M }}{r}\right)\left[1+2k_1P_2(\cos\theta)+2\left(1-\frac{2{ M}}{r}\right)^{-1}\frac{J^{2}}{r^{4}}(2\cos^2\theta-1)\right]dt^2
\\+\left(1-\frac{2{M}}{r}\right)^{-1}\left[1-2k_2P_2(\cos\theta)-2\left(1-\frac{2{M}}{r}\right)^{-1}\frac{J^{2}}{r^4}\right]dr^2
\\+r^2[1-2k_3P_2(\cos\theta)](d\theta^2+\sin^2\theta d\phi^2)-4\frac{J}{r}\sin^2\theta dt d\phi\,
\end{split}
\end{equation}
where
\begin{eqnarray}\label{ht2}
k_1&=&\frac{J^{2}}{{M}r^3}\left(1+\frac{{M}}{r}\right)-\frac{5}{8}\frac{Q-J^{2}/{M}}{{M}^3}Q_2^2\left(\frac{r}{{M}}-1\right), \qquad\qquad\qquad k_2=k_1-\frac{6J^{2}}{r^4}, \nonumber\\
k_3&=&k_1+\frac{J^{2}}{r^4}-\frac{5}{4}\frac{Q-J^{2}/{M}}{{M}^2r}\left(1-\frac{2{M}}{r}\right)^{-1/2}Q_2^1\left(\frac{r}{M}-1\right),\qquad P_{2}(x)=\frac{1}{2}(3x^{2}-1),\nonumber\\
Q_{2}^{1}(x)&=&(x^{2}-1)^{1/2}\left[\frac{3x}{2}\ln\frac{x+1}{x-1}-\frac{3x^{2}-2}{x^{2}-1}\right],\qquad\qquad\qquad\qquad Q_{2}^{2}(x)=(x^{2}-1)\left[\frac{3}{2}\ln\frac{x+1}{x-1}-\frac{3x^{3}-5x}{(x^{2}-1)^2}\right].\nonumber
\end{eqnarray}
Here $P_{2}(x)$ is Legendre polynomials of the first kind, $Q_l^m$ are the associated Legendre polynomials of the second kind and the constants ${M}$, ${J}$ and ${Q}$ are the total mass, angular momentum and  quadrupole parameter of a rotating star respectively. 
The Hartle-Thorne metric is an approximate solution of vacuum Einstein field equations that describes the exterior of any slowly and rigidly rotating, stationary and axially symmetric body. The metric is given with accuracy up to the second order terms in the body's angular momentum, and first order in its quadrupole moment. 
The approximate Kerr metric \cite{Kerr} in the Boyer-Lindquist coordinates $(t,\ R,\ \Theta,\ \phi)$ up to second order terms in the rotation parameter $a$ can be obtained from (\ref{ht1}) by setting
\begin{equation}\label{tr1}
J=-Ma,\quad Q={J}^2/{M},
\end{equation}
and making a coordinate transformation given by
\begin{eqnarray}\label{tr2}
r&=&R+\frac{a^2}{2R}\left[\left(1+\frac{2M}{R}\right)\left(1-\frac{M}{R}\right)-\cos^2\Theta\left(1-\frac{2M}{R}\right)\left(1+\frac{3M}{R}\right)\right], \\
\theta&=&\Theta+\frac{a^2}{2R^2}\left(1+\frac{2M}{R}\right)\sin\Theta\cos\Theta. \nonumber\
\end{eqnarray}

\section{The Epicyclic Frequencies}

Properties of congruences of nearly circular geodesic orbits in stationary and axially symmetric spacetime such as the HT spacetime are studied because of their fundamental role in the theory of accretion disks around compact objects with strong gravity. The radial epicyclic frequency  and vertical epicyclic frequency are the most important characteristics of these orbits. Analytic formulas for the frequencies in Schwarzschild, Kerr and Hartle-Thorne metrics have been published many times by several authors (see e.g. \cite{wald1984,okazaki1987,perez1997,A2003,abr2004}) and are well-known. Here we give more explicit derivation of the frequencies for the HT metric at equatorial plane. 

The Keplerian angular velocity (angular frequency) $\omega_{K}$ for the Hartle-Thorne solution is given by
\begin{equation}
\omega_{K}^2(u)=\omega_{K0}^2(u)\left[1\mp j F_{1}(u)+j^2 F_{2}(u)+q F_{3}(u)\right]
\end{equation}
where $(+/-)$ stands for co-rotating/contra-rotating geodesics, $j=J/M^2$ and $q=Q/M^3$ are the dimensionless angular momentum and quadrupole parameter and $u=M/r$. The rest quantities are defined as follows
\begin{eqnarray}
\vspace{-1cm} 
\omega_{K0}^2(u)&=&u^{3}/M^2,\quad F_{1}(u)=2u^{3/2}, \nonumber \\
F_{2}(u)&=&\frac{48u^7-80u^6+12u^5+26u^4-40u^3-10u^2-15u+15}{8u^2(1-2u)}-F(u),\qquad\qquad\qquad\qquad\nonumber \\
F_{3}(u)&=&-\frac{5(6u^4-8u^3-2u^2-3u+3)}{8u^2(1-2u)}+F(u),\nonumber \\
F(u)&=&\frac{15(1-2u^3)}{16u^3}\ln\left(\frac{1}{1-2u}\right).\nonumber 
\end{eqnarray}

Let us consider oscillations around circular orbits in the plane
$\theta=\pi/2$. As $U_t=-E$ and $U_\phi=L$ are the integrals of motion, related to the energy and the orbital angular momentum of a test boby, then the perturbation of 4-velocity becomes
\begin{equation}
\delta U_{\alpha}=(-\delta E, \delta U_r(t), \delta U_\theta(t), \delta L).
\end{equation}
To calculate the epicyclic frequencies of these oscillations we can use the 4-velocity on the circular orbit
\begin{equation}
g^{\alpha\beta}U_\alpha U_\beta=\epsilon,
\end{equation}
where $\epsilon=0,-1$, and calculate it on the perturbed trajectory
\begin{gather}
g^{\alpha\beta}(r+\delta r,\theta+\delta\theta)(U_\alpha+\delta U_\alpha) (U_\beta+\delta U_\beta)=\epsilon,\\
\left(\delta r\frac{\partial g^{\alpha\beta}}{\partial r}+\delta\theta\frac{\partial g^{\alpha\beta}}{\partial \theta}+ \frac{\delta r^2}{2}\frac{\partial^2 g^{\alpha\beta}}{\partial r^2}+ \frac{\delta\theta^2}{2}\frac{\partial^2 g^{\alpha\beta}}{\partial \theta^2}+ \delta r\delta \theta\frac{\partial^2 g^{\alpha\beta}}{\partial r\partial\theta}+\dots\right)U_\alpha U_\beta\nonumber\\
+2g^{\alpha\beta}(r,\theta)\delta U_\alpha U_\beta+g^{\alpha\beta}(r,\theta)\delta U_\alpha \delta U_\beta=0.
\end{gather}

As $\theta=\pi/2$ is the plane of symmetry, then all derivatives with $\partial\theta$ are equal to 0 (but not with $\partial\theta^2$). Equation of circular orbits can be written as
\begin{equation}
g^{\alpha\beta}_{,r}U_\alpha U_\beta=0,
\end{equation}
where $g^{\alpha\beta}_{,r}$ is the derivative of the metric tensor with respect to $r$.
As a result first nonvanishing terms give us
\begin{equation}
\left(\frac{\delta r^2}{2}\frac{\partial^2 g^{\alpha\beta}}{\partial r^2}+\frac{\delta \theta^2}{2}\frac{\partial^2g^{\alpha\beta}}{\partial\theta^2}\right) U_\alpha U_\beta+(U^t)^2\left[g_{rr}\dot{\delta r}^2+g_{\theta\theta}\dot{\delta\theta}^2\right]=C
\end{equation}
where the dot denotes derivative with respect to $t$ and
\begin{equation}
C=2g^{t\phi}(E\delta L+L\delta E+\delta E\delta L)-g^{tt}(\delta E^2+2E\delta E)-g^{\phi\phi}(\delta L^2+2L\delta L),
\end{equation}
is a constant. Explicitly, using the fact that $U_\alpha=(-E,0,0,L)$ is
constant, we have after division by $E^2$
\begin{equation}\label{eqeq}
\left(\frac{\delta r^2}{2}\frac{\partial^2}{\partial r^2}+ \frac{\delta\theta^2}{2}\frac{\partial^2}{\partial\theta^2}\right)(g^{tt}-2g^{t\phi}l+g^{\phi\phi}l^2)+\left(g^{t\phi}l-g^{tt}\right)^2\left[g_{rr}\dot{\delta r}^2+g_{\theta\theta}\dot{\delta\theta}^2\right]=C/E^2,
\end{equation}
where $l=-U_{\phi}/U_{t}$ is the specific orbital angular momentum of a test particle per unit energy.
Taking derivative of (\ref{eqeq}) with respect to $t$ one has the equation of non-coupled harmonic oscillations with angular frequencies
given by
\begin{equation}
\omega_{x}^2=
\frac{\partial_{xx}(g^{tt}-2g^{t\phi}l+g^{\phi\phi}l^2)}{
2g_{xx}\left(g^{t\phi}l-g^{tt}\right)^2}
,\quad x=r\text{ or }\theta.
\end{equation}

Defining an effective potential by: $U_{eff}(r,\theta,l)=
g^{tt}-2lg^{t\phi}+l^2g^{\phi\phi}$, which can be used to find the general formula for the epicyclic frequencies on the equatorial plane, the previous expression becomes
\begin{equation}
\omega_{x}^{2}=\frac{(g_{tt}+\omega_{K} g_{t\phi})^2}{2g_{xx}}\left(\frac{\partial^{2}U_{eff}}{\partial x^2}\right)_{l}, \quad\quad  x \quad\epsilon\quad (r, \theta)
\end{equation}
where $\omega_{K}=U^{\phi}/U^{t}$ is the orbital angular velocity of a test particle.

The radial frequencies are given by
\begin{eqnarray}
\omega_{r}^{2}(u)& =&\omega_{r0}^{2}(u)\left[1\pm j X_{1}(u)+j^2X_{2}(u)+qX_{3}(u)\right],
\end{eqnarray}
where
\begin{align}
\omega_{r0}^2(u)&=u^3(1-6u)/M^2,\qquad X_{1}(u)=\frac{6u^{3/2}(1+2u)}{(1-6u)},\\
X_{2}(u)&=X(u)-\frac{384u^8-720u^7-112u^6+404u^5+162u^4+130u^3-635u^2+375u-60}{8u^2(1-2u)(1-6u)},\\
X_{3}(u)&=\frac{5(48u^5+30u^4+26u^3-127u^2+75u-12)}{8u^2(1-2u)(1-6u)}-X(u),\\
X(u)&=\frac{15(1-2u)(2u^2+13u-4)}{16u^3(1-6u)}\ln\left(\frac{1}{1-2u}\right),
\end{align}
The vertical frequencies are given by
\begin{eqnarray}
\omega_{\theta}^{2}(u)& =&\omega_{\theta0}^{2}(u)\left[1\mp j Y_{1}(u)+j^2Y_{2}(u)+q Y_{3}(u)\right],
\end{eqnarray}
where
\begin{align}
\omega_{\theta0}^2(r)&=u^3/M^2,\qquad Y_{1}(u)=6u^{3/2},\\
Y_{2}(u)&=\frac{48u^7-224u^6+28u^5+66u^4+170u^3-295u^2+165u-30}{8u^2(1-2u)}+Y(u),\qquad\qquad\quad\\
Y_{3}(u)&=-\frac{5(6u^4+34u^3-59u^2+33u-6)}{8u^2(1-2u)}-Y(u),\\
Y(u)&=\frac{15(2-u)(1-2u)^2}{16u^3}\ln\left(\frac{1}{1-2u}\right).
\end{align}


\section{Determination of the Mass, Angular Momentum and Quadrupole Moment}
Spacetimes around rotating NSs can be with a high precision approximated by the three parametric Hartle--Thorne (HT) solution of Einstein field equations (\cite{har-tho:1968}; see \cite{ber-etal:2005}). The solution considers the mass $M$, angular momentum $J$ and quadrupole moment $Q$ (supposed to reflect the rotationally induced oblateness of the star).  It is known that in most situations modeled with the present NS equations of state (EoS) the NS external geometry is very different from the Kerr geometry (representing the limit of HT geometry for $\tilde{q}\equiv QM/J^2\rightarrow1$). However, the situation changes when the NS mass approaches maximum for a given EoS. For high masses the quadrupole moment does not induce large differences from the Kerr geometry since $\tilde q$ takes values close to unity. Nevertheless this does not mean that one can easily neglect the quadrupole moment. For this reason in this work we extend the analyses of \cite{ste-vie:1999} involving the Hartle-Thorne solution. 

Usually in the literature the QPOs data are given by the following frequencies
\begin{eqnarray}
f_{\phi}(u)&=&\omega_{K}(u)/(2\pi),\qquad f_{r}(u)=\omega_{r}(u)/(2\pi),\qquad f_{\theta}(u)=\omega_{\theta}(u)/(2\pi).
\end{eqnarray}

\begin{figure}[t]
\includegraphics[angle=0,totalheight=6.5cm,width=8.5cm]{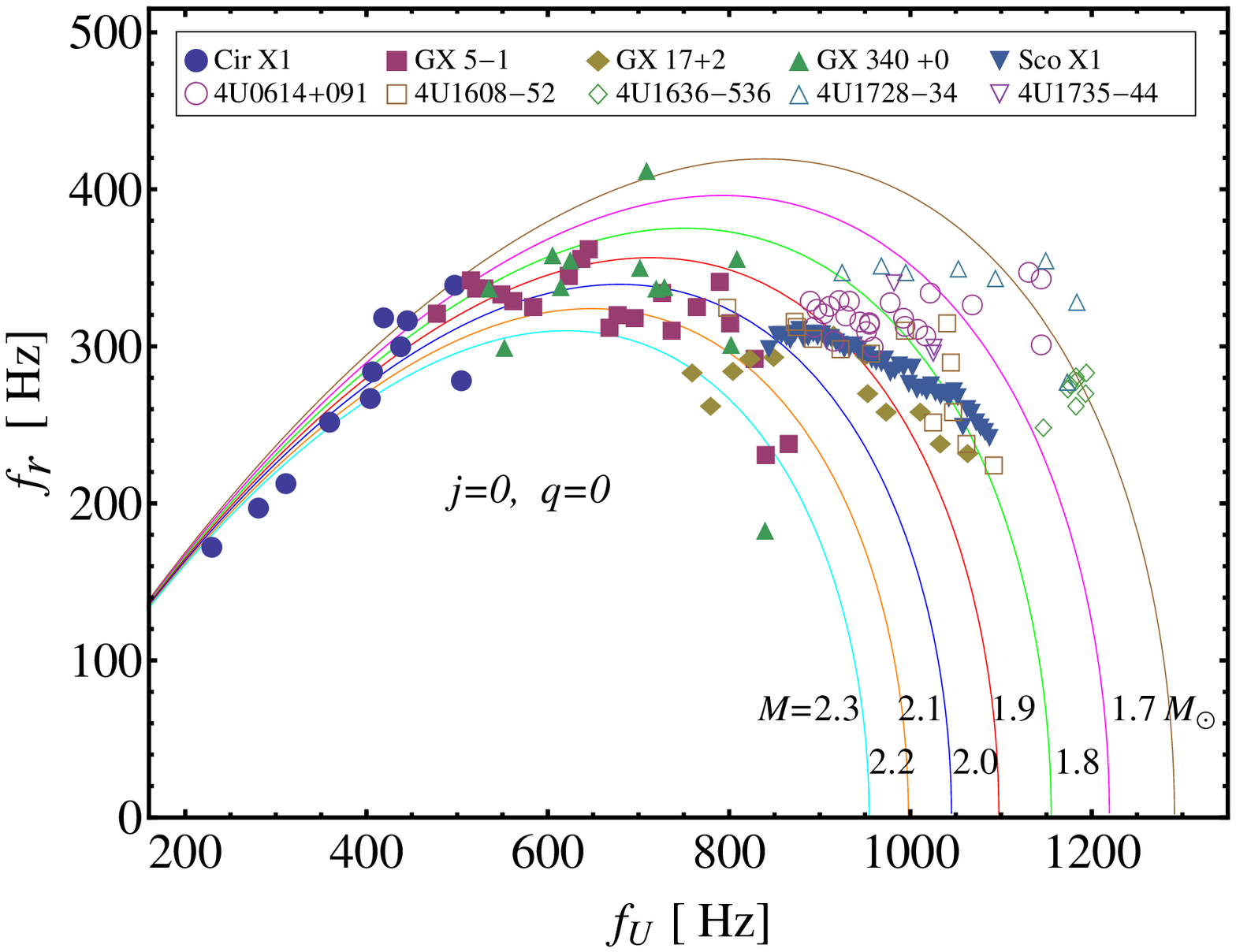}\hfill
\includegraphics[angle=0,totalheight=6.5cm,width=8.5cm]{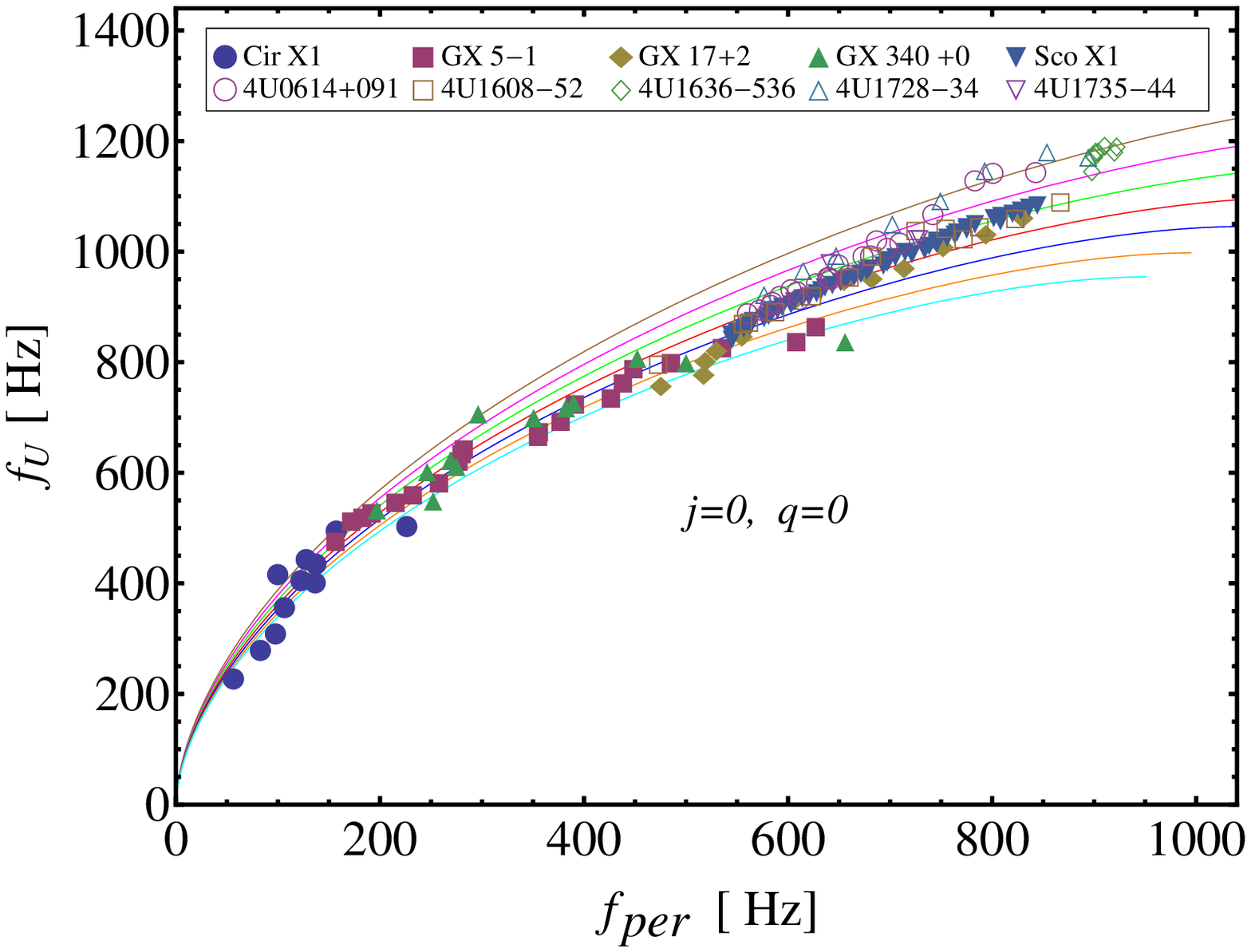}\\
\parbox[t]{0.47\textwidth}{\caption{The radial frequency $f_r$ is plotted versus the upper frequency  $f_\phi$ for the Schwarzschild spacetime ($j=0,\, q=0$). From top to bottom $M=[1.7,\ 1.8,\ 1.9,\ 2.0,\ 2.1,\ 2.2,\ 2.3,\ 2.4]M_{\odot}$.}\label{plotM}}\hfill
\parbox[t]{0.47\textwidth}{\caption{The upper frequency $f_\phi$ is plotted versus the periastron frequency  $f_{per}=f_\phi-f_r$ for the Schwarzschild spacetime ($j=0,\, q=0$). From top to bottom the mass $M=[1.7,\ 1.8,\ 1.9,\ 2.0,\ 2.1,\ 2.2,\ 2.3,\ 24]M_{\odot}$.}\label{plotMper}}
\end{figure}

\begin{figure}[t]
\includegraphics[angle=0,totalheight=6.5cm,width=8.5cm]{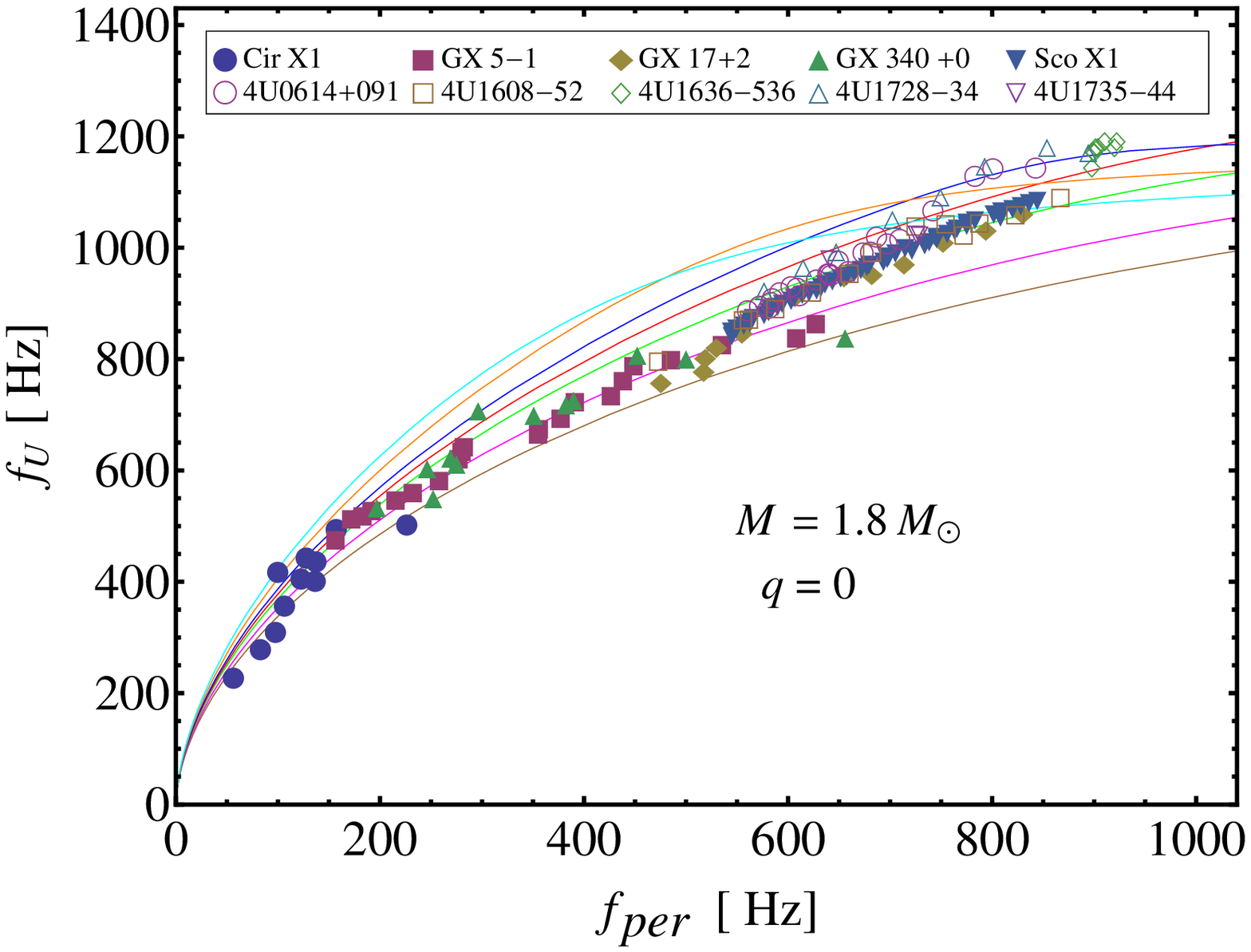}\hfill
\includegraphics[angle=0,totalheight=6.5cm,width=8.5cm]{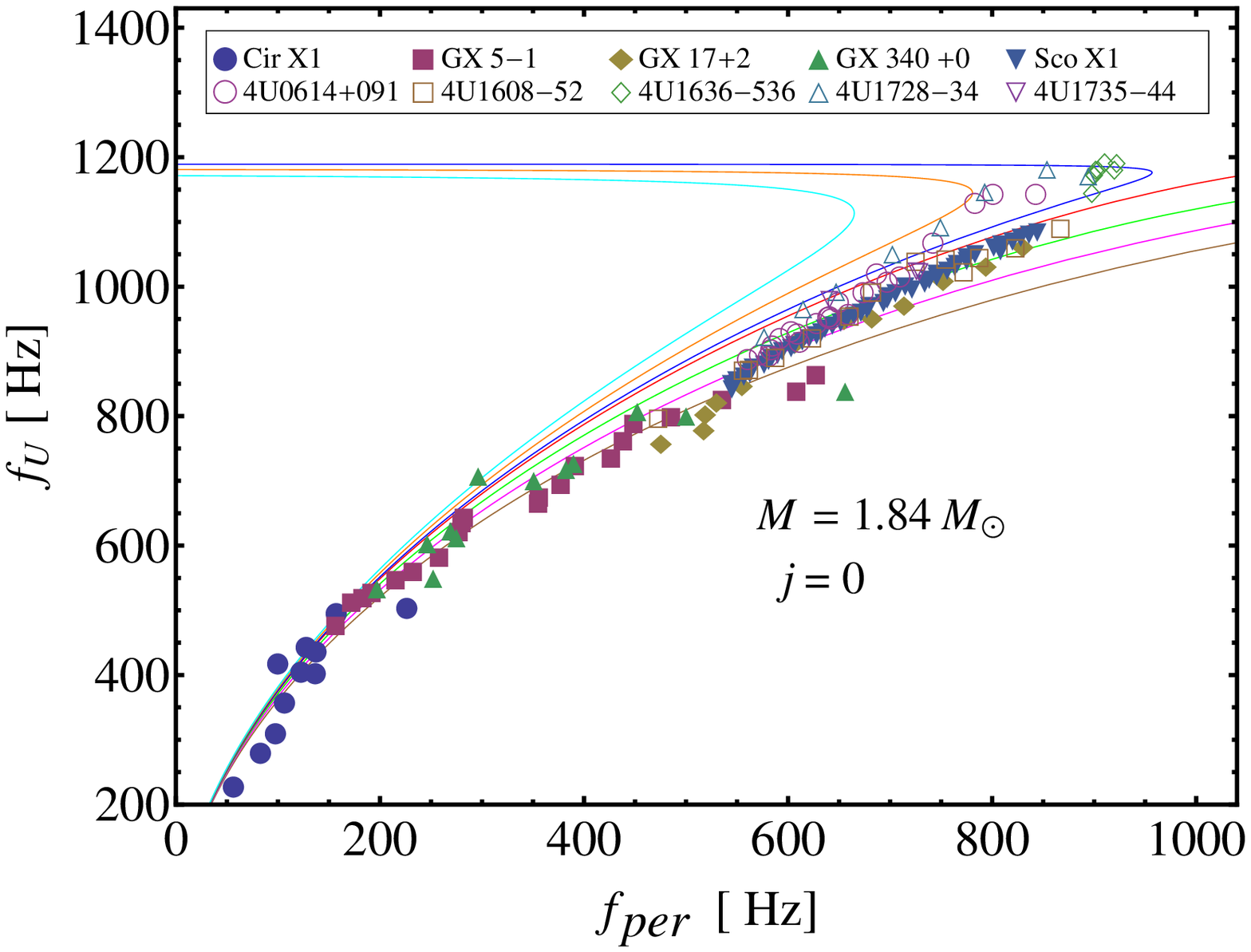}\\
\parbox[t]{0.47\textwidth}{\caption{The upper frequency $f_U$ is plotted versus the periastron frequency  $f_{per}$ for the fixed  $M=1.8 M_{\odot}$ and $q=0$. From bottom to top (or from right to left) $j=[-0.5,\, -0.3,\, -0.1,\, 0,\, 0.1,\, 0. 3,\, 0.5]$.}\label{plotJper}}\hfill
\parbox[t]{0.47\textwidth}{\caption{The upper frequency $f_U$ is plotted versus the periastron frequency  $f_{per}$ for the fixed  $M=1.84 M_{\odot}$ and $j=0$. From bottom to top (or from right to left) $q=[-1.2,\, -0.7,\, -0.3,\, 0,\, 0.1,\, 0.3,\, 0.5]$.}\label{plotQper}}
\end{figure}

\begin{figure}[t]
\includegraphics[angle=0,totalheight=6cm,width=8.5cm]{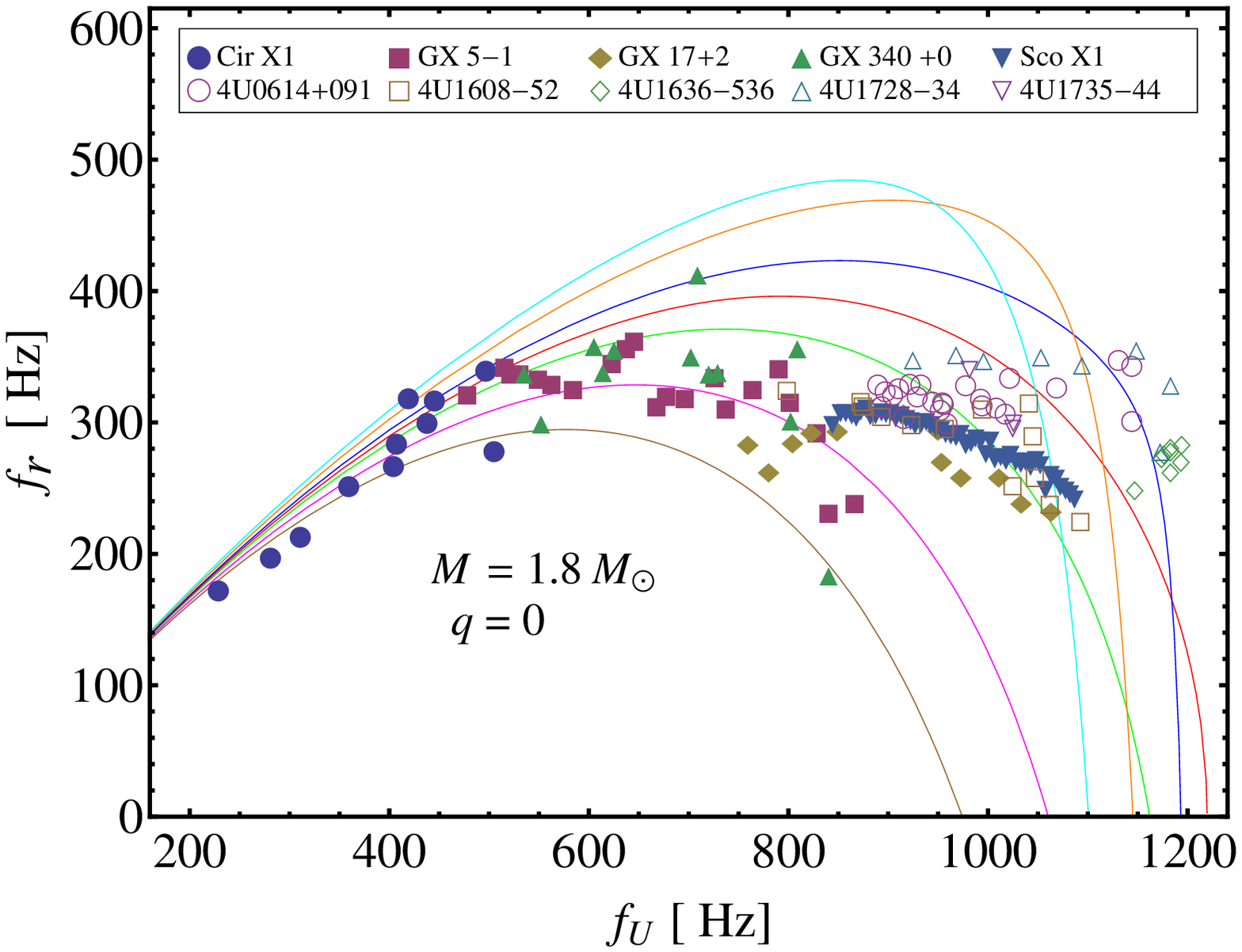}\hfill
\includegraphics[angle=0,totalheight=6cm,width=8.5cm]{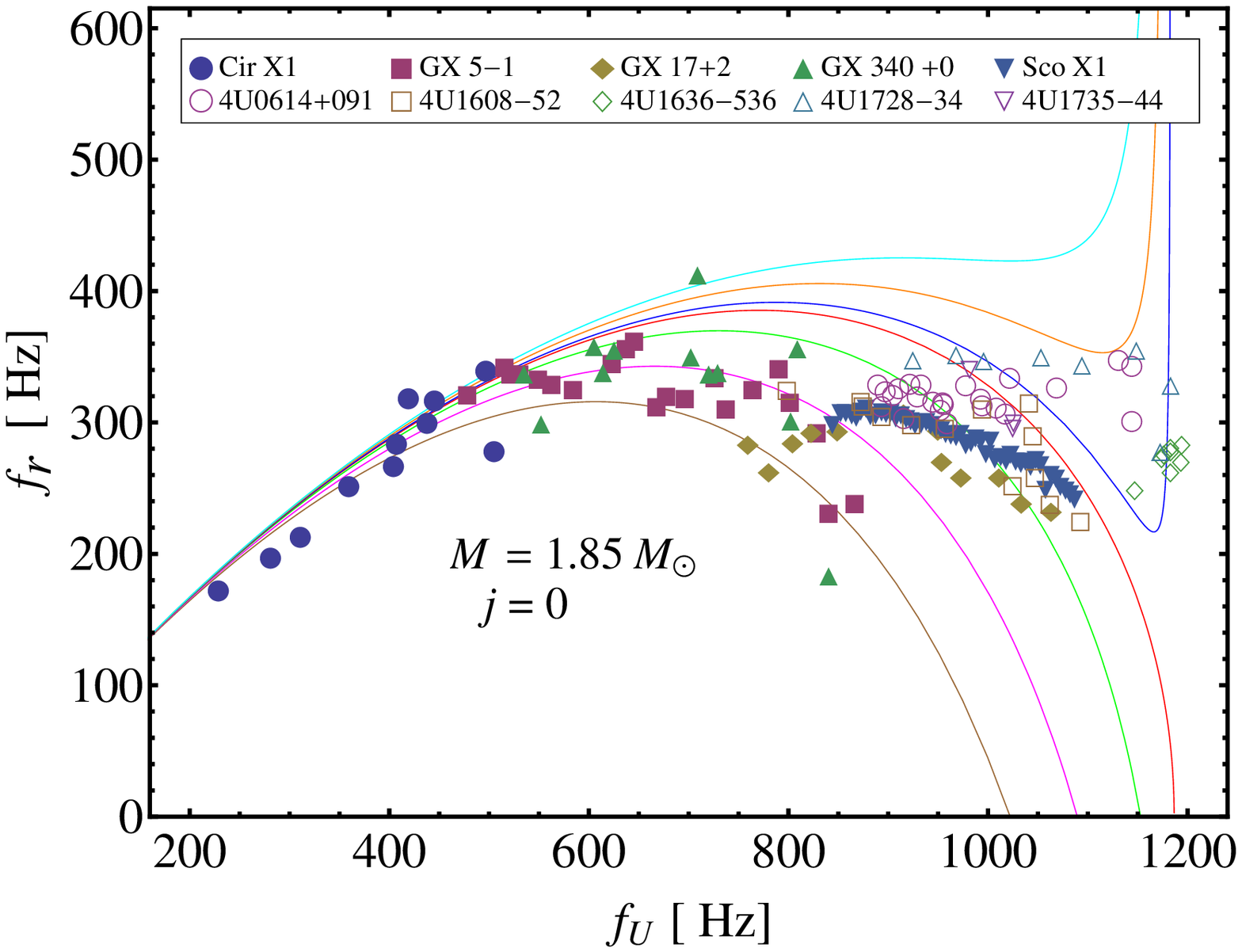}\\
\parbox[t]{0.47\textwidth}{\caption{The radial frequency $f_r$ is plotted versus the upper frequency  $f_U$ for the fixed $M=1.8 M_{\odot}$ and $q=0$. From bottom to top $j=[-0.5,\, - 0.3,\, -0.1,\, 0,\, 0.1,\, 0.3,\, 0.5]$.}\label{plotJ}}\hfill
\parbox[t]{0.47\textwidth}{\caption{The radial frequency $f_r$ is plotted versus the upper frequency  $f_U$ for the fixed $M=1.85 M_{\odot}$ and $j=0$. From bottom to top $q=[-2,\, -1,\,- 0.3,\, 0,\, 0.1,\, 0.3,\, 0.5]$.}\label{plotQ}}
\end{figure}

\begin{figure}[t]
\includegraphics[angle=0,totalheight=6cm,width=8.5cm]{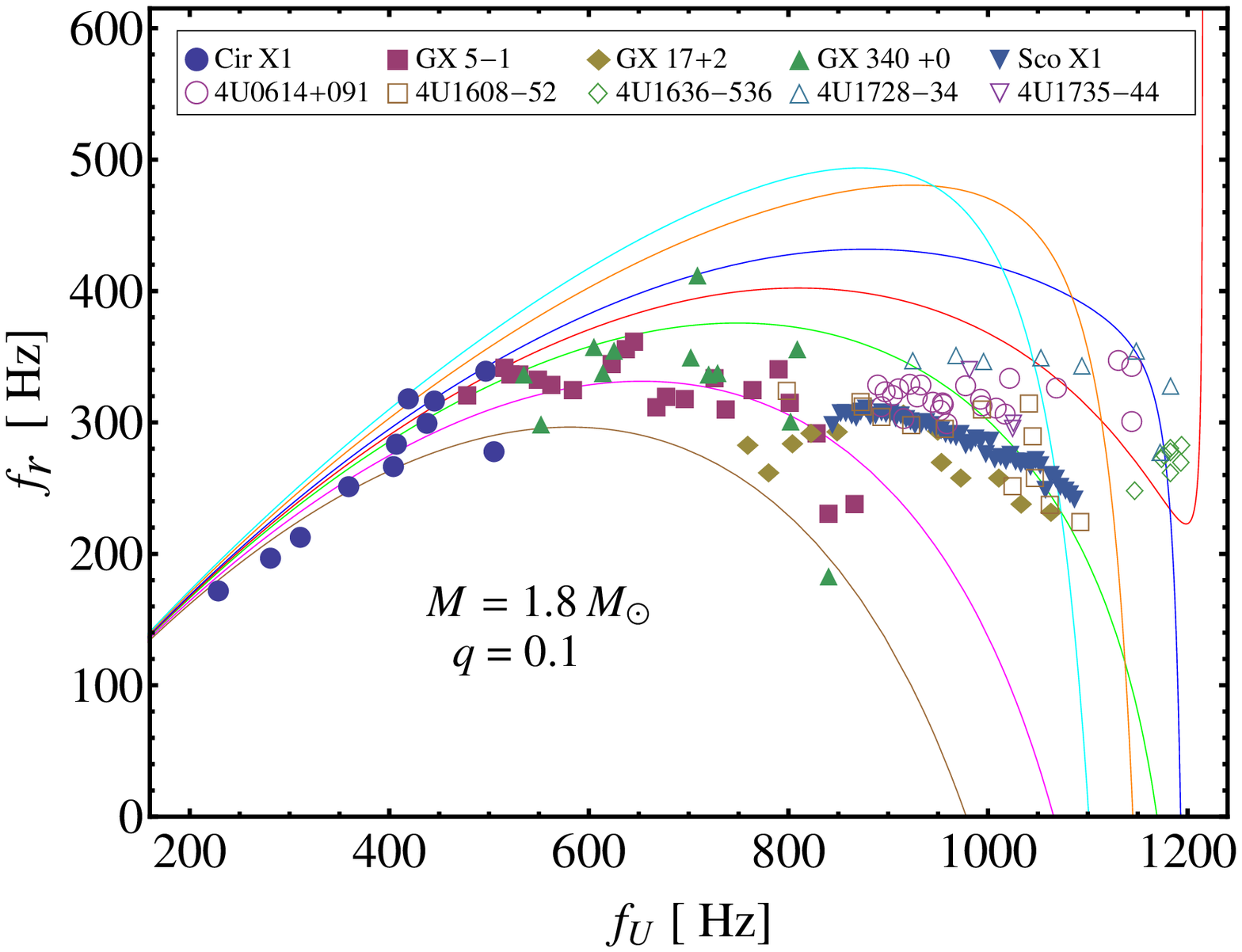}\hfill
\includegraphics[angle=0,totalheight=6cm,width=8.5cm]{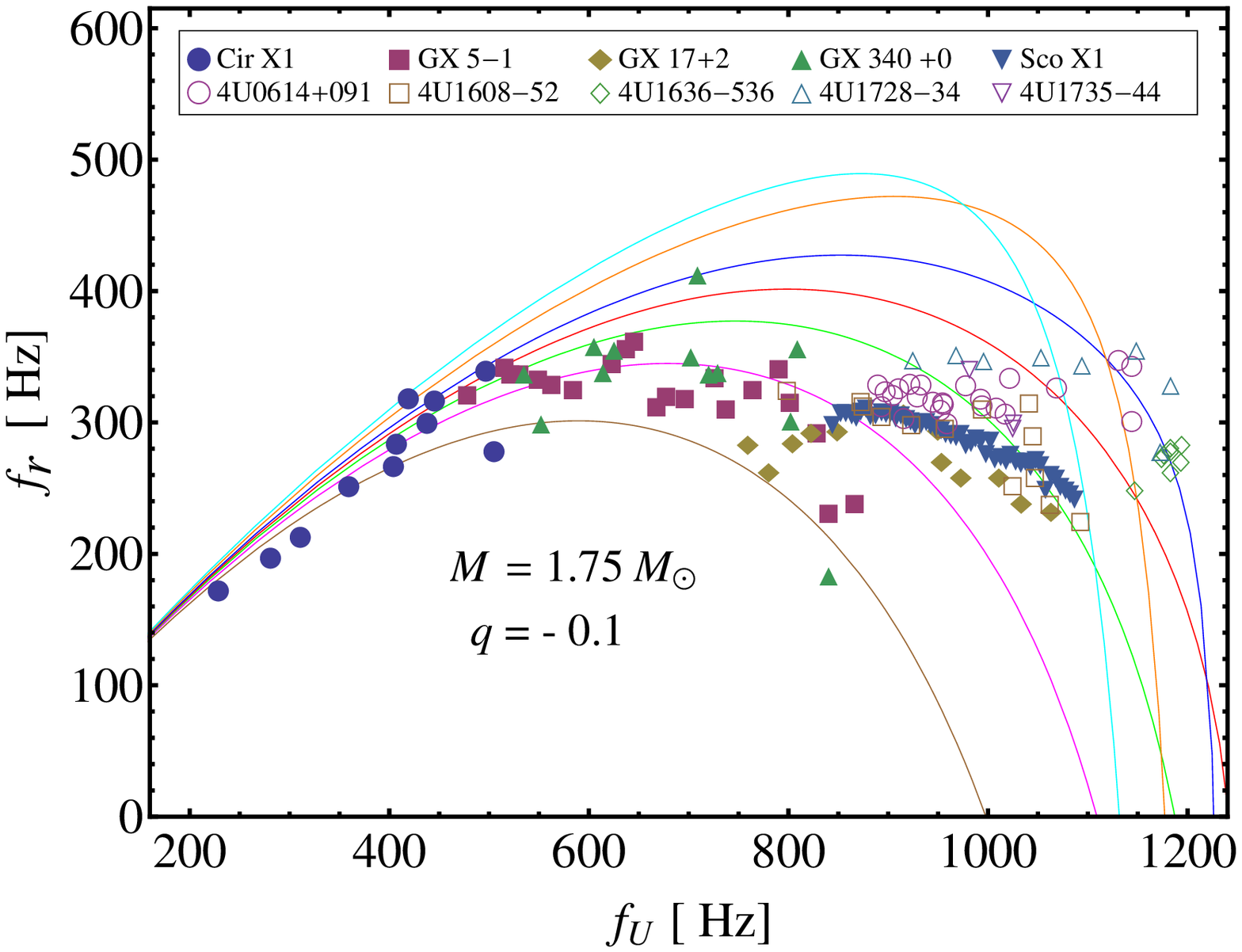}\\
\parbox[t]{0.47\textwidth}{\caption{The radial frequency $f_r$ is plotted versus the upper frequency  $f_U$ for the fixed $M=1.8 M_{\odot}$ and $q=0.1$. From bottom to top $j=[-0.5,\, -0.3,\, -0.1,\, 0,\, 0.1,\, 0.3,\, 0.5]$.}\label{plotJpQ}}\hfill
\parbox[t]{0.47\textwidth}{\caption{The radial frequency $f_r$ is plotted versus the upper frequency  $f_U$ for the fixed $M=1.75 M_{\odot}$ and $q=-0.1$. From bottom to top $j=[-0.5,\, -0.25,\, -0.1,\, 0,\, 0.1,\, 0.3,\, 0.5]$.}\label{plotJmQ}}
\end{figure}

\begin{figure}[t]
\includegraphics[angle=0,totalheight=6cm,width=8.5cm]{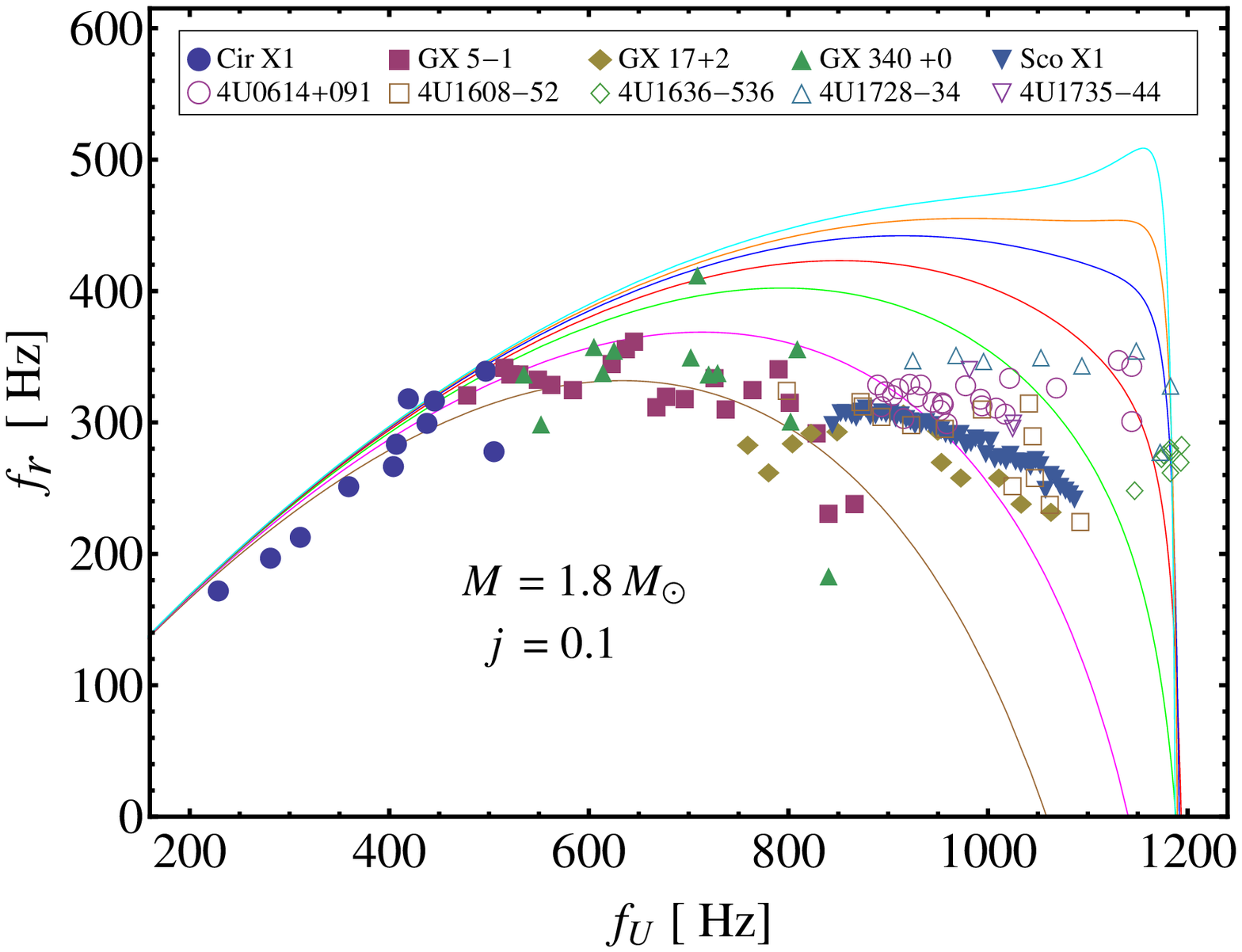}\hfill
\includegraphics[angle=0,totalheight=6cm,width=8.5cm]{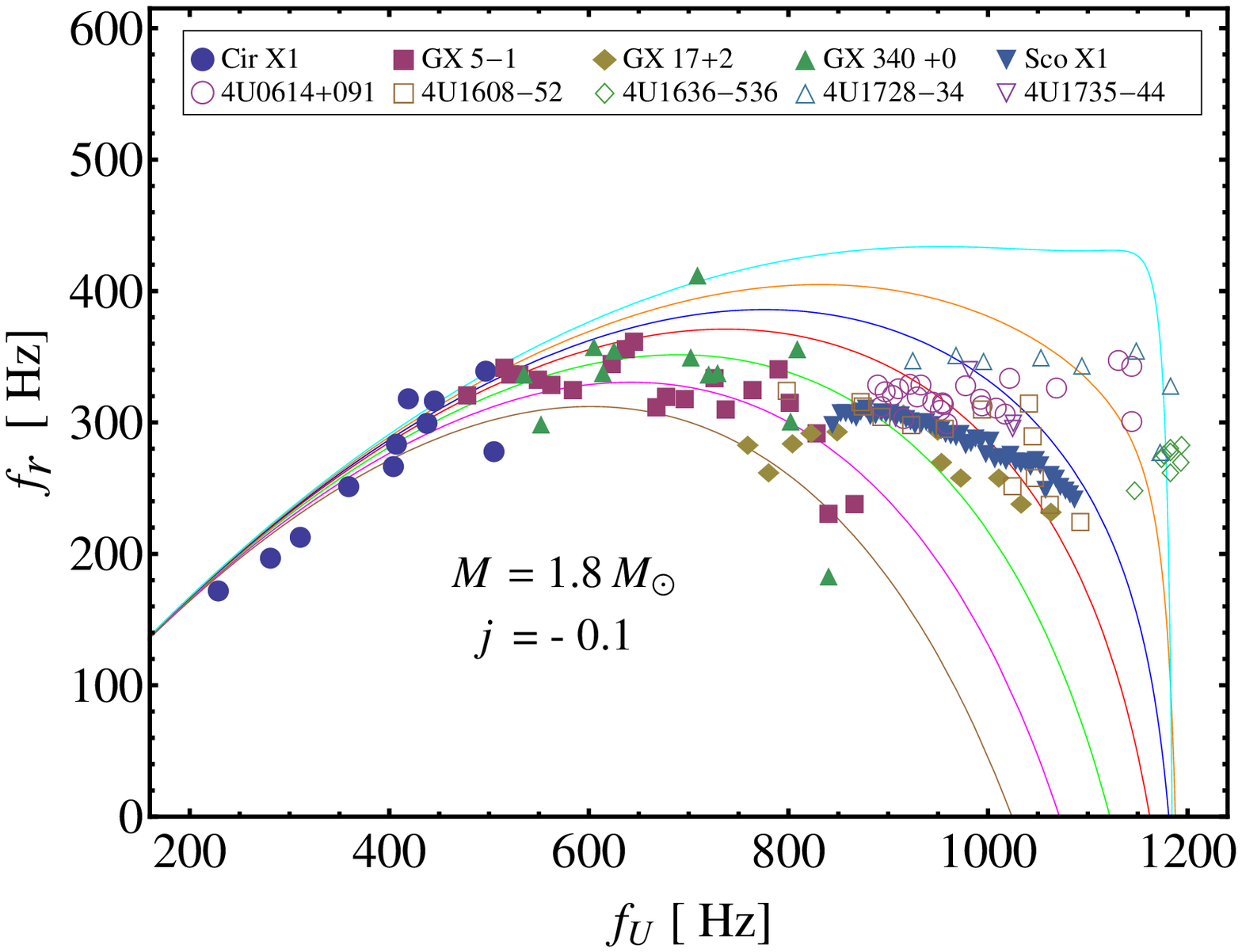}\\
\parbox[t]{0.47\textwidth}{\caption{The radial frequency $f_r$ is plotted versus the upper frequency  $f_U$ for the fixed  $M=1.8 M_{\odot}$ and $j=0.1$. From bottom to top $q=[-2.2,\, -1,\, -0.3,\, 0,\, 0.2,\, 0.3,\, 0.4]$.}\label{plotQpJ}}\hfill
\parbox[t]{0.47\textwidth}{\caption{The radial frequency $f_r$ is plotted versus the upper frequency  $f_U$ for the fixed  $M=1.8 M_{\odot}$ and $j=-0.1$. From bottom to top $q=[-2,\, -1.2,\, -0.5,\, 0,\, 0.3,\, 0.6,\, 0.9]$.}\label{plotQmJ}}
\end{figure}

In Fig.~\ref{plotM} the radial frequency $f_{r}$ is plotted versus the upper (Keplerian) frequency $f_U$ in the Schwarzschild spacetime ($j=0,\ q=0$). The smaller mass the higher radial frequency $f_{r}$. Note for ($j=0,\ q=0$), the Keplerian frequency $f_K$ coincides with the vertical frequency $f_{\theta}$. The Keplerian frequency $f_{\phi}$ versus the periastron frequency $f_{per}=f_{\phi}-f_r$ is show in Fig. \ref{plotMper}. The observational datapoints in the Figs. \ref{plotM}, \ref{plotMper} belong to Atoll ( 4U0614+091,  4U1608-52, 4U1636-536, 4U1728-34, 4U1735-44) and Z ( GX 5-1, GX 17+2, GX 340+0, Sco X-1, Cir X-1) sources.  For the sake of clarity the error bars have been omitted. The QPOs data have been taken from \cite{mendez1998, mendez1999, mendez2000, mendez2007} and references therein. For the fixed mass and zero quadrupole parameter different curves are shown in Figs. \ref{plotJper} and \ref{plotJ} varying the value of $j$. Figs.~\ref{plotQper} and \ref{plotQ} show the frequencies around non-rotating deformed objects with the fixed mass and different quadrupole moment. Indeed, the rotation and deformation of a central object (in our case a neutron stars) play a pivotal role in describing quasi-periodic oscillations. In Figs. \ref{plotJpQ}, \ref{plotJmQ}, \ref{plotQpJ}, \ref{plotQmJ} are shown the radial frequencies versus the upper frequencies depending on $M$, $j$ and $q$ of the central source.

In this work we used the minimum set of parameters such as the total mass $M$, dimensionless angular momentum $j$ and quadrupole parameter $q$ of the source. Unfortunately, from the observations it is hard to obtain precise values of the masses of the LMXBs. Different references show contradicting numbers. For example, Sco X-1 is the well-known X-ray binary system classified as a low-mass X-ray binary; the neutron star is roughly 1.4 solar masses, while the donor star is only 0.42 solar masses \cite{Steeghs2002}. However references \cite{ste-vie:1999},\cite{Zhang2001} and \cite{muk:2009} present various values for the neutron star mass.

\begin{figure}[t]
\includegraphics[angle=0,totalheight=9cm,width=13cm]{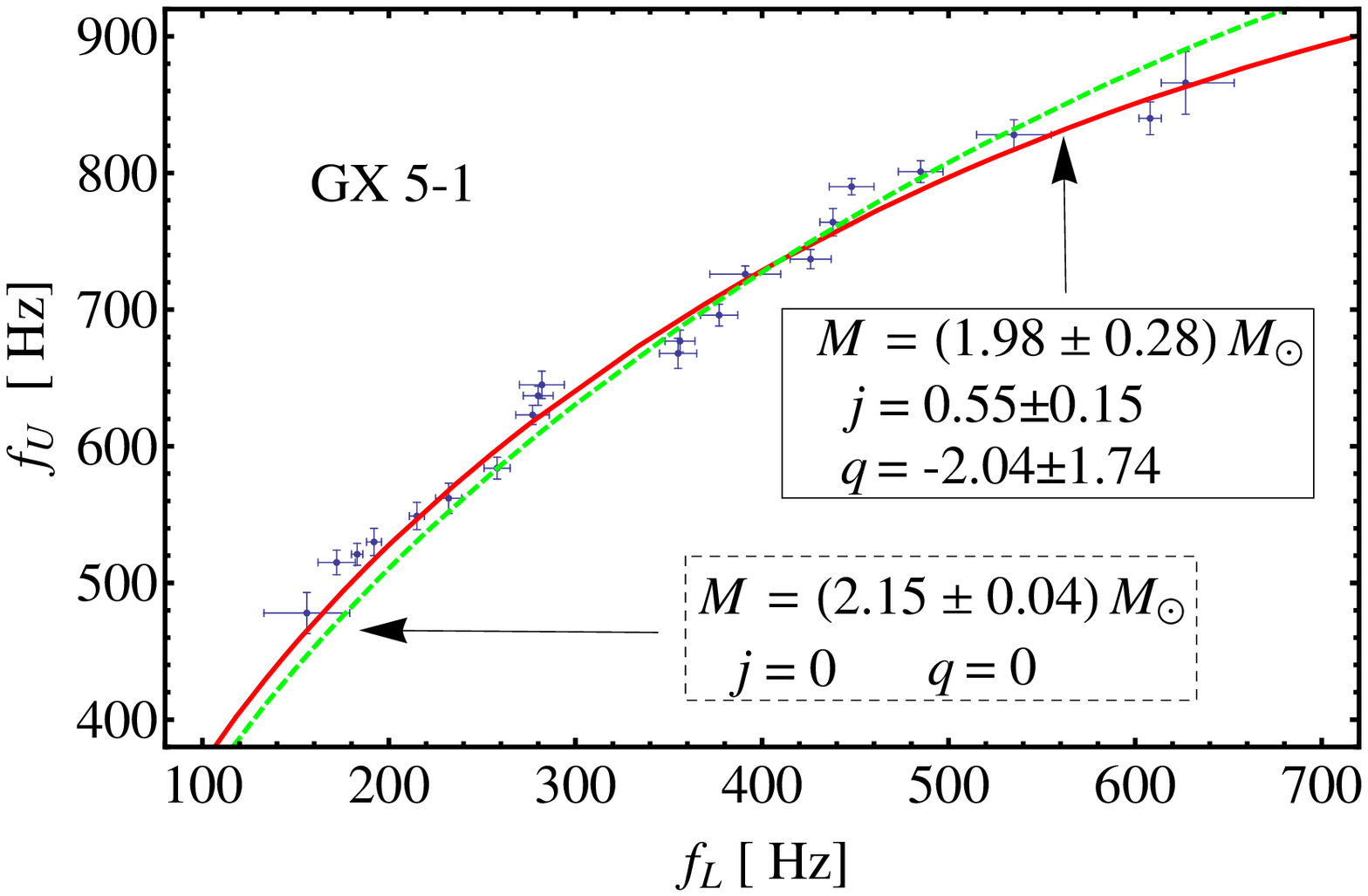}
\caption{The upper frequency $f_U$ is plotted versus the lower frequency  $f_L$ for the Z source GX 5-1. The dashed green line corresponds to the static case and the solid red line corresponds to the rotating case.}\label{fig:gx51}
\end{figure}

In Fig.~\ref{fig:gx51} we show best fits for the upper frequency versus the lower frequency for the Z source GX 5-1. Here we see that the fit with all three parameters is better than the fit with the only one parameter, the mass $M$. Statistical test $\chi^2=0.998$ for the three parameter fit and  $\chi^2=0.993$ for the one parameter fit.


\section{Conclusion}
In this paper we have derived the formulas for the epicyclic frequencies of test particles in the Hartle-Thorne spacetime. With the help of these frequencies and according to the relativistic precession model we have interpreted the quasi-periodic oscillations of the low-mass X-ray binaries. We have constructed the dependence of the higher frequencies with respect to the lower frequencies varying the main parameters of the central compact object such as the mass, angular momentum and quadrupole moment. Eventually for Z source GX 5-1 we have performed the fitting analyses and found the best fit to estimate the mass, angular momentum and quadrupole moment. We have shown that the three parameter fit is better than one-two parameter fit.
For better analyses one needs to consider more sources with refined data. It would be interesting to perform further calculations assigning the neutron star equation of state and construct mass-radius, mass-angular momentum, angular velocity-quadrupole moment etc. relations in order to compare and contrast the theory with observations. A first step in this direction was recently taken forward by Belvedere et. al. \cite{belvedere}, where the authors construct the equilibrium configurations of uniformly rotating neutron stars taking into account all fundamental interactions in the self-consistent relativistic fashion. The construction of the angular momentum-quadrupole moment relation from the theoretical point of view and its comparison with the best fit of observational data will be the issue of future investigations.

\section*{Acknowledgements}
The authors thank ICRANet for support. One of them (K.B.) thanks Mariano M\'endez for providing observational data and for stimulating discussions.

\end{document}